# Global Features of Nucleus-Nucleus Collisions in the Ultrarelativistic Domain

M.V. Savina, S.V. Shmatov, N.V. Slavin, P.I. Zarubin,

*Joint Institute for Nuclear Research,*

10 September, 1998

## Abstract

HIJING generator simulation of nucleus-nucleus collisions at ultrarelativistic energies are presented. Is shown, that the global characteristics of nucleus-nucleus collisions, such as of distribution of a charge multiplicity, total and electromagnetic transverse energy over pceudorapidity are rather sensitive to some predictions of formation models of high-exited nuclear medium, namely parton energy losses in dense nuclear matter. This losses results to appearance of a broad maximum in global variable distributions over pseudorapidity. The most profound of this effect occurs at central heavy ion collisions at LHC energy.

Global characteristics of nucleus-nucleus collisions available in the CMS experiment [1] are differential distributions of total transverse energy flows, $dE_t/d\eta$, in a pseudorapity range $-5 \leq \eta \leq 5$ as well as electromagnetic and charged multiplicity ones, $dE_t^{\gamma}/d\eta$ and $dN_{ch}/d\eta$. For instance, the CMS calorimeter system will be able to cover about 80 % of a total transverse energy flow. Use of the CMS apparatus for studies of global observable distributions allow one to establish a quite general rules of nucleus-nucleus collision dynamics up to the highest energy frontier with a huge rapidity coverage and to verify some crucial predictions of quark-gluon plasma formation models in a sufficiently simple way.

The test signal of formation of a dense nuclear matter in high-energy ion collisions could be J/ψ and dijet supression, charm and strange particles production induced by thermal emissions [2]. One of the predicted features of such state of a nuclear matter is energy losses of scattered partons in final state interactions with a dense nuclear matter called jet quenching [3]. The radiative energy losses of high-energy partons in nuclear matter equal:

$$\frac{dE}{dx} \sim \alpha_s \mu_D^2 \ln^2(E/\mu_d),$$

where $\alpha_s$ is a running constant, $\mu_D$ is Debye screening lengths of the medium. So, this losses depen-dens on the matter density and can be used as sensitivity probe of the hot chromodynamic matter formation.

The HIJING Monte Carlo generator developed by M. Gyulassy and X.-N. Wang [4,5] provide an opportunity to study various manifestations of this mechanism. Among other effects originated by a jet quenching one may expect a significant modification in the differential distributions mentioned above. Indeed a strong indication is found on appearance of a wide bump in the interval $-2 \leq \eta \leq 2$ over a pseudorapidity plateau of such distributions for the case when jet quenching is switched on [6]. Note that a space-time picture of parton shower evolution is not considered in the HIJING, and the effect of dense matter formation is introduced by a phenomenological way.

The problem of existence of this bump rise the question of principal importance - whether a asymptotic behavior of basic distributions established already at lower energy scale will be broken in new energy domain or not. A wide pseudorapidity plateau provided by the LHC machine and the CMS acceptance give a chance to obtain a definite answer to the question.

Even in absence of such modifications a conservation of similarity in a behavior of global observable distributions with respect to lower energy ones make it possible to derive a conclusion on color transparency of colliding nuclear matter. This circumstance is of a special meaning since already ≈ 80% of total transverse energy flow in LHC energy scale are calculated in the framework of application of pQCD to the HIJING [5].

So, a well defined choice between two possible scenarios make these distributions interesting for dedicated experimental study and for definition of future explorations. Below We shall present most important results of our study.

In addition to a general physics interest the global event characteristics provide a basis for estimation of a collision impact parameter and inelasticity characterization for correlated measurements in other reaction channels like jet and lepton pair production. For instance this measurement might provide a basis for searches of manifestation of special variation of nuclear structure functions [7]. Besides, in the framework of global energy simulations one can work out practical suggestions on a fast inelastic event selection with various collision centralities in order to enrich most central event sample (≈ 3 %).

In this respect it's interesting to note that such secondary effect as a jet quenching can modify only the central rapidity part leaving both fragmentation regions ($3 \leq |\eta| \leq 5$) unchanged. This circumstance is useful for definition of nuclear collision geometry (i.e. impact parameter) by means of the very forward calorimeters with minimal dependence on collision dynamics details in the central region (fig. 1).

Let's turn back to results of the global energy flow simulations. The HIJING generator was used to produce large samples of events (up to 10,000) of minimal bias PbPb, NbNb, CaCa, OO, αα, pp interactions. In fig. 2 differential transverse energy distributions $dE_t/d\eta$ are presented for Pb-Pb collisions at energy values $\sqrt{S_{nn}}$ equal to 5, 3, 1, 0.5, 0.2, 0.1 TeV/nucleon (c.m.). A distinctive feature of the distributions is the presence of the bump in central pseudorapidity ($-2 \leq \eta \leq 2$) region starting from the energy value larger than 3 TeV/nucleon. For lower energy values this effect is not so profound due to smaller rapidity difference between fragmentation regions. Verification of such dependence impose a task of feasibility of colliding beam energy scanning with the LHC machine from few hundred GeV up to few TeV region.

The existing CERN heavy ion injection system can provide sufficient luminosity for whole variety of fully stripped nuclei with rapid transition from one type ion to another [8]. This makes it possible to study a mass number dependence of the distribution modifications. Reduction of the colliding nucleus radii must lead to a bump reduction. This provides an additional test of jet quenching existence. Fig. 3 shows that lead-lead collisions demonstrate maximum quenching dependence while in lighter ion cases the bump becomes less and less profound. It's interesting to note that a degree of shadowing modification of PDF's can be estimated by measurement of absolute value of transverse energy flow for various ion species.

Obviously, that the energy losses are proportional to a mean path traversing by parton in the medium: $\Delta E(l)=ldE/dx$. Therefore, these losses will be sensitive to the sizes of space occupied a high-exited nuclear substance, formed starting at critical energy density about 1-2 GeV/fm. Such energy density are achievement at a maximum overlap of the colliding nuclei, i.e. mainly in central nucleus-nucleus collisions. In this connection it would be important to be able to estimate as strongly this effect depends from a centrality of an event and as it is reflected in the shape of distributions of global variables on pceudorapidity. A collision impact parameter dependence was followed for a PbPb case from most central collisions to peripheral ones (Fig. 4). It was shown that a bump stay distinguishable until parameter values about 12 fm. So, this gives additional option to study effect by a global approach making impact parameter tagging by the very forward calorimeter. As it follows from Fig. 2 fluctuations in nucleus-nucleus collisions of an impact parameter definition at a level of 2 fm which is sufficient to explore transverse energy correlation's over an available pseudorapidity range on event-by-event basis.

Parton energy losses in dense nuclear matter lead to an increase of the particle production at small and moderate $p_t$ and to a decrease of it at large $p_t$ simultaneously. It is possible to expect, that the jet quenching effect will also reflecting in distributions of a multiplicity over pceudorapidity. The distribution of the charge multiplicity over pseudorapidity for case of PbPb collisions with and without jet quenching are shown on Fig.5a. It was found in the framework of this study that the qualitatively the same picture of a bump formation are obtained in differential distributions of transverse electromagnetic energy (Fig.5b) and charged multiplicity flow. This gives experimental alternatives to search for the effect.

Our preliminary results indicate that the qualitatively same modification of a distribution shape can be obtained in the VENUS Monte Carlo model [9] in the case when rescattering of nucleons is switched on. It's appears to be very fruitful to compare the discussed distributions with predictions of other models and their options.

Let's discuss some practical requests needed to be resolved for the global observable studies with CMS. First of them is availability of heavy ion beams for the CMS at LHC during period when the CMS solenoid isn't switched yet. This circumstance is critical to allow one to measure undistorted distributions of total energy and charged multiplicity flows. Heavy ion option might be attractive of alignment of the tracking detectors. Otherwise it's necessary to prove that for a case of gamma-quantum energy flow a distortion produced in the electromagnetic calorimeter by charged hadron flux is small enough.

Another practical problem is sufficiency of CMS calorimeter resolution for a bump observation. One of the CMS calorimeter practical problems is energy resolution depletion on boundaries of the major calorimeter part, i.e. barrel, forward, and very forward ones. This problem can be overcome in case nucleus-nucleus collision by normalization of the distributions to proton-proton ones having the same dependence of calorimeter responses. It is shown in the paper [10] that application of this idea allow one to observe the bump with the CMS calorimeter resolution even on a sample of 100 PbPb events without complicated corrections.

So, it might be concluded that the CMS experiment with a wide calorimeter and tracker (pixel part) acceptance is able to observe the described modification of global observable distributions. From experimental point of view this option to search for jet quenching is specially attractive due to the positive correlation - than more jet quenching is than more transverse energy "repumped" toward $\eta = 0$. In fact this leads to effective increase of energy density or "stopping power" in the central region.

These differential distributions can be obtained with a practically unlimited accuracy for various colliding nuclei by a direct counting of differential energy and charged multiplicity flow. The special demand of this study is availability of a brief period of CMS running with the solenoid switched off and, as addition, variation of an LHC collision energy value and a type of accelerated nuclei. By these reasons global observable measurements of nucleus-nucleus collisions in ultrarelativistic domain demanding minimum beam time might pretend to become a candidate on "Physics of LHC first seconds".

We would like to express our thanks to Profs. I.A. Golutvin and A.I. Malakhov for encouraging support and stimulating discussions.

# Figure captions

Fig. 1. Correlation between total energy flow per collision (vertical axis in GeV; full scale 100,000 GeV) in very forward calorimeter direction ($3 \leq |\eta| \leq 5$) and collision impact parameter (horizontal axis in fm). HIJING Model. From top to bottom: 10,000 PbPb, NbNb, CaCa collisions at 5 TeV/nucleon (c.m.).

Fig. 2. Differential distribution of total transverse energy $dE_t/d\eta$ (GeV) over pseudorapidity for PbPb collisions at energy values $\sqrt{S_{nn}}$ equal to 5, 3, 1, 0.5, 0.2, 0.1 TeV/nucleon (c.m.). Jet quenching is on. Normalized per number of events; $\eta$ bin size is 0.087.

Fig. 3. Differential distribution of total transverse energy flow $dE_t/d\eta$ (GeV) over pseudorapidity for PbPb, NbNb, CaCa, OO, αα collisions at 5 TeV/nucleon (c.m.) with and without jet quenching. Normalized per number of events; $\eta$ bin size is 0.087.

Fig. 4. A collision impact parameter dependence for a PbPb case ions from $b \leq 2$ fm up to $b \leq 20$ fm.

Fig. 5. Charged multiplicity $dN_{ch}/d\eta$ (a) and electromagnetic transverse energy $dE_t^{\gamma}/d\eta$ (b) distributions over pseudorapidity for PbPb, NbNb, CaCa, OO, αα collisions at 5 TeV/nucleon (c.m.) with and without jet quenching. Normalized per number of events; $\eta$ bin size is 0.087.

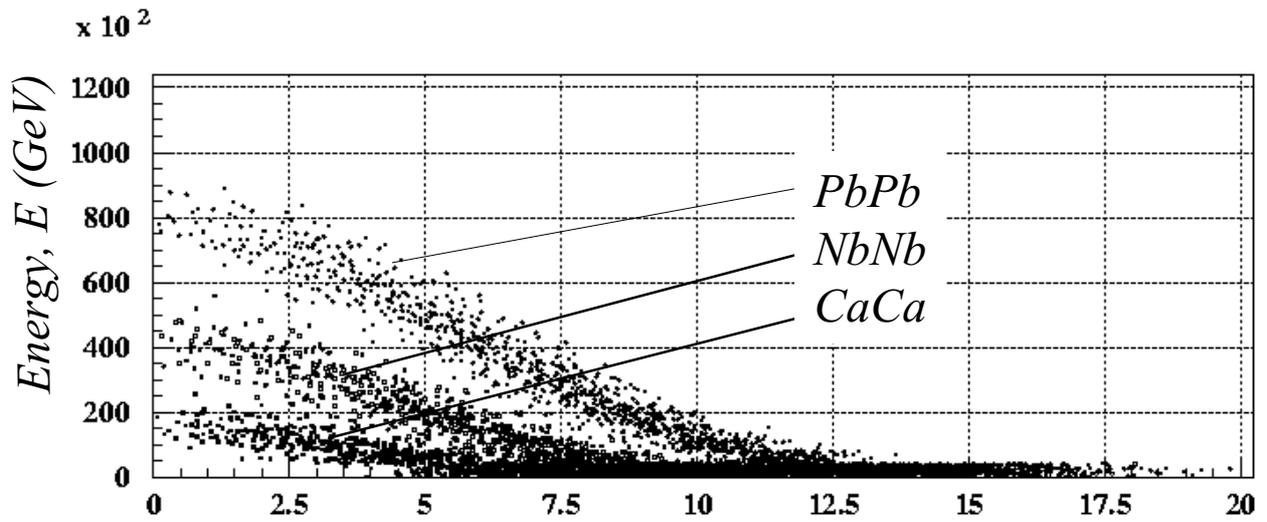
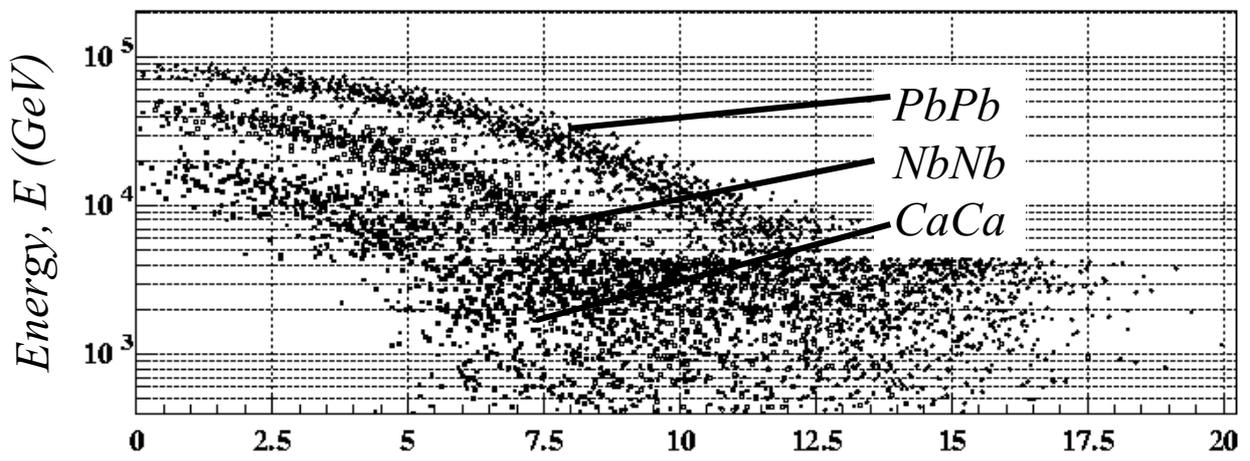

Impact parameter, $b(fm)$

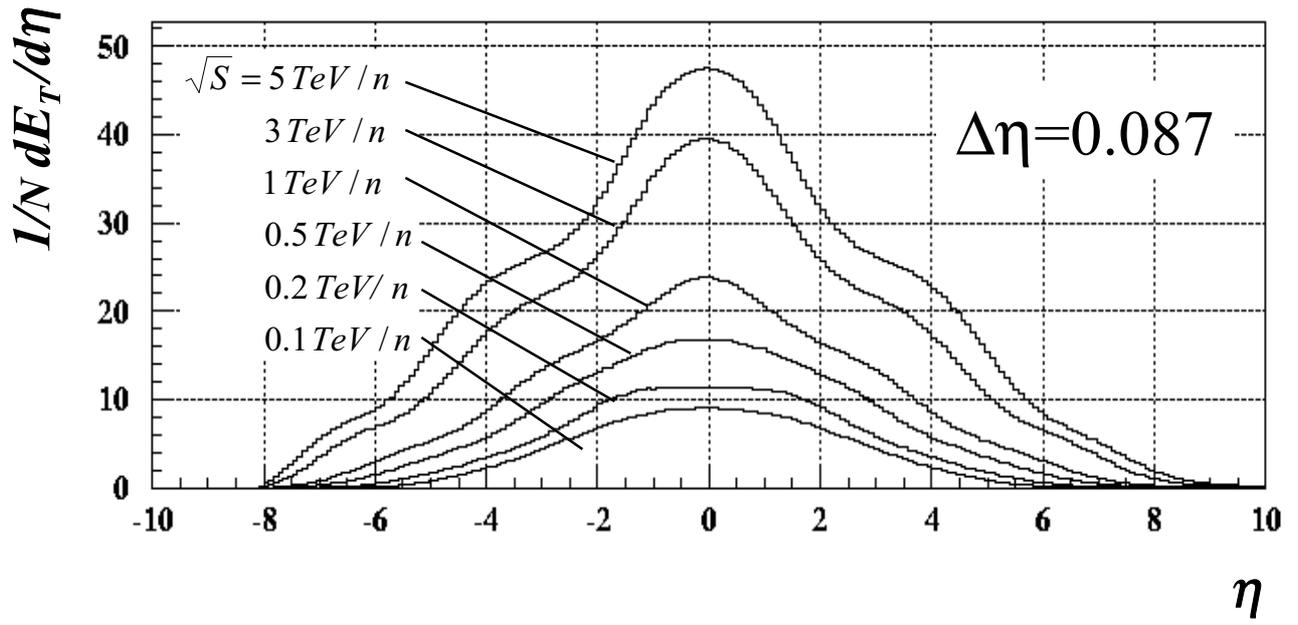

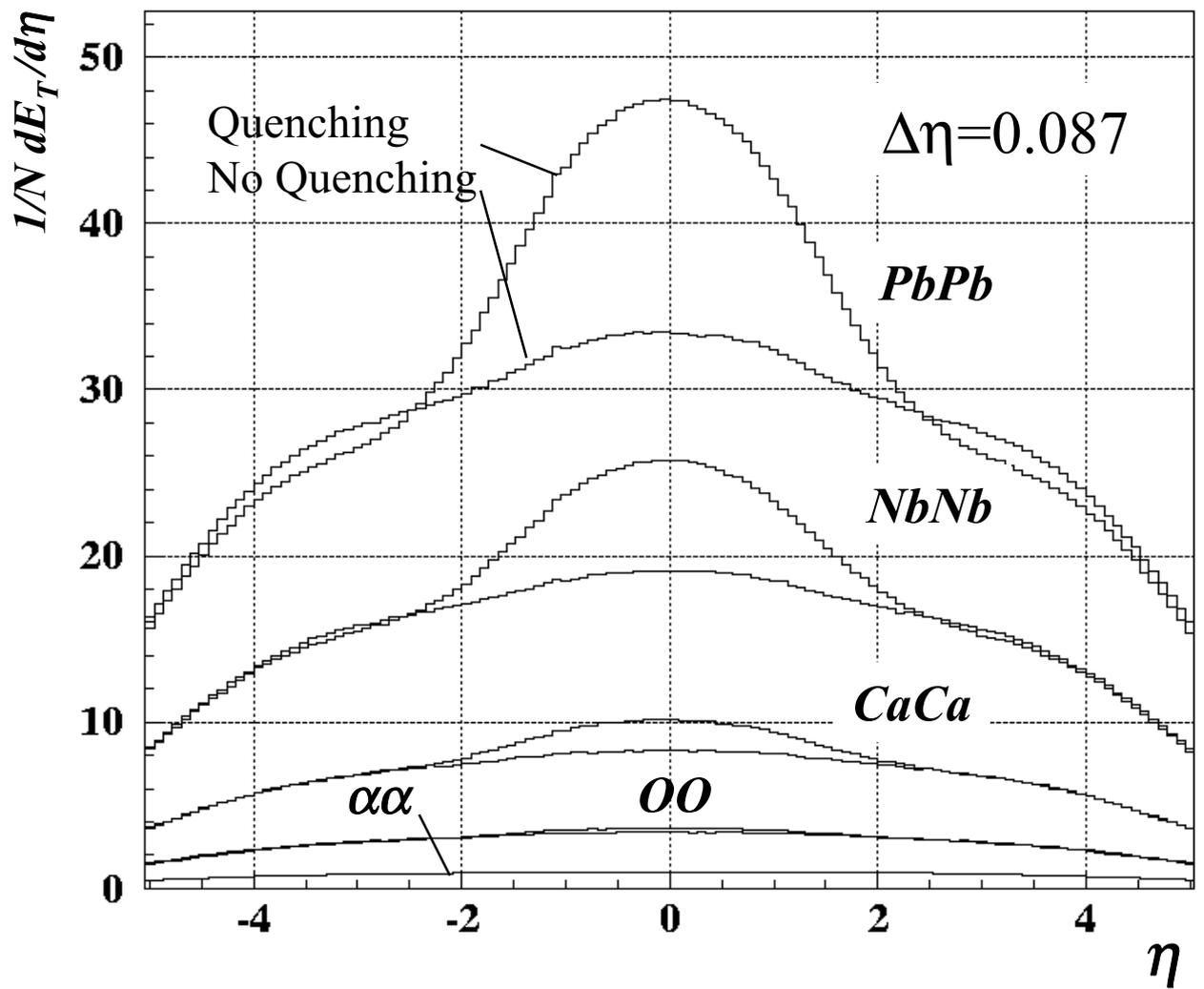

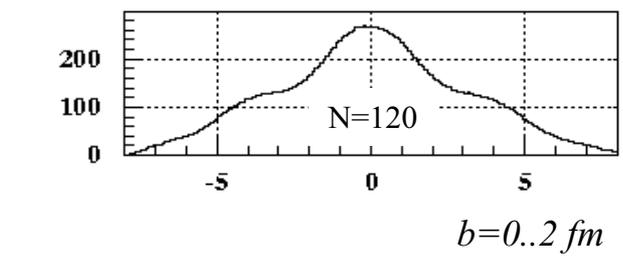

*b=0..2 fm*          *b=4..6 fm*

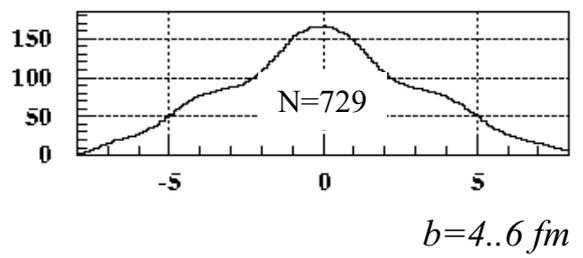

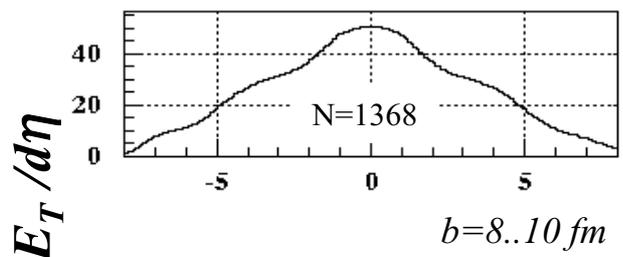

*b=8..10 fm*         *b=10..12 fm*

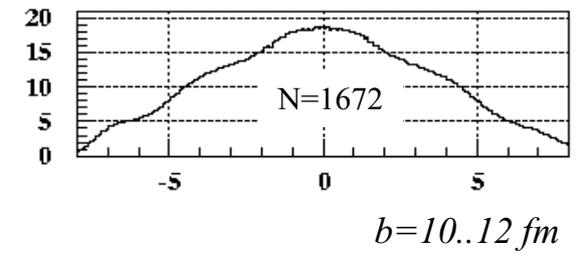

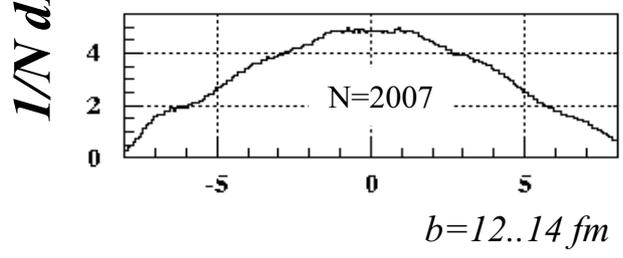

*b=12..14 fm*        *b=14..16 fm*

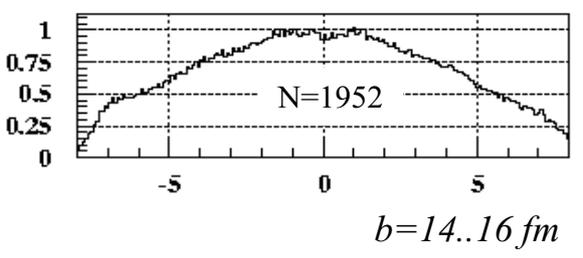

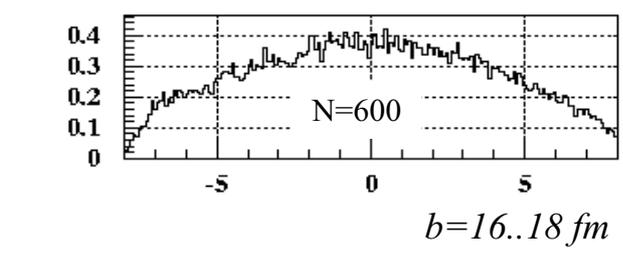

*b=16..18 fm*        *b=18..20 fm*

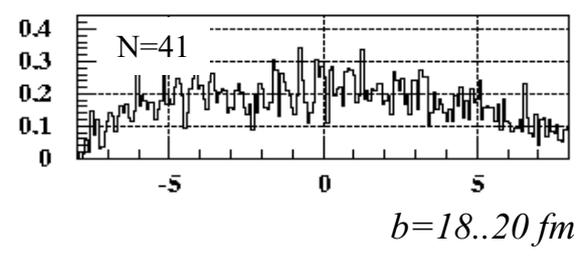

$\eta$

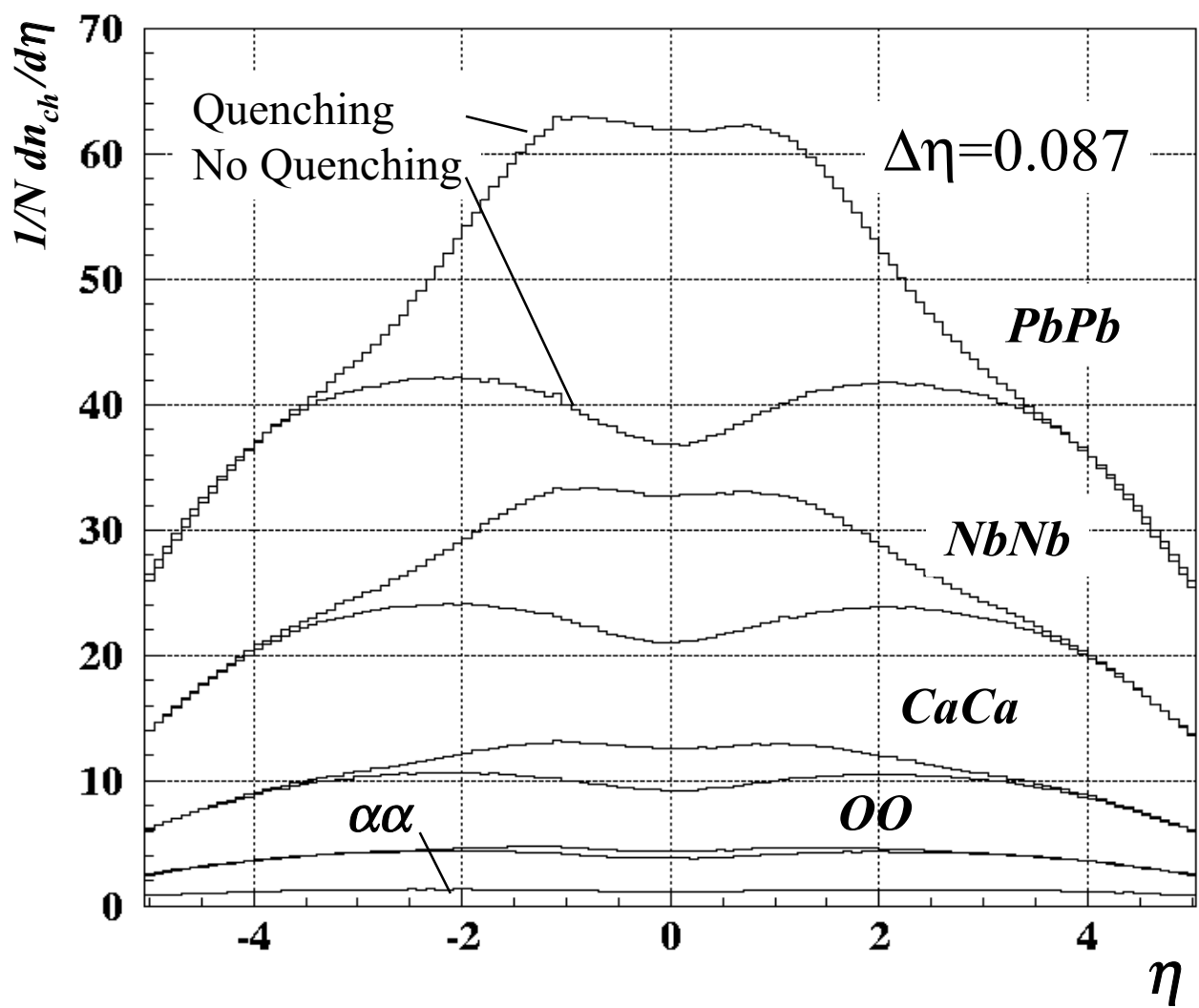